\documentclass[12pt]{article}
\usepackage[left=2.5cm,right=2.5cm,top=2.5cm,bottom=2.5cm]{geometry}
\usepackage[T1]{fontenc}
\usepackage{fix-cm}  
\usepackage{graphicx}  
\usepackage{float}  
\usepackage[left]{lineno}  
\usepackage{indentfirst}  
\usepackage{xcolor}  
\usepackage{setspace}
\usepackage{siunitx}
\usepackage[
    backend=biber,
    style=nature,
    sorting=none,
    isbn=false,
    clearlang=true,
    url=false,
    hyperref=true,
    doi=true,
    autocite=superscript
]{biblatex}  
\usepackage[
    figurename=Fig.,
    font=small,
    labelfont=bf,
    labelsep=period
]{caption}  
\usepackage{amsmath}
\usepackage{amsfonts}
\usepackage{amssymb}
\usepackage{upgreek}
\usepackage[
    bookmarksnumbered=true,
    colorlinks=true,
    linkcolor=blue,
    anchorcolor=.,
    citecolor=blue,
    filecolor=.,
    menucolor=.,
    runcolor=.,
    urlcolor=blue,
    linktoc=all
]{hyperref}  
\usepackage{newtxmath,newtxtext}  

\usepackage[separate-uncertainty=true]{siunitx}

\newenvironment{sciabstract}{%
\begin{quote} \bf}
{\end{quote}}

\DeclareUnicodeCharacter{2212}{\(-\)}
\title{Metallicity-driven polar transitions in \\ topological epilayers}
\author{Eduardo D. Stefanato$^{1,*}$, Nicolas M. Kawahala$^{1,*}$, Bianca A. Kawata$^2$\\
Paulo H. O. Rappl$^2$, Eduardo Abramof$^2$, Felix G. G. Hernandez$^{1,\dagger}$\\
\\
\normalsize{$^1$Instituto de Física, Universidade de São Paulo, São Paulo, SP 05508-090, Brazil.}\\
\normalsize{$^2$National Institute for Space Research, São José dos Campos, SP 12201-970, Brazil.}\\
\\
\normalsize{$^*$These authors contributed equally to this work.}\\
\normalsize{$^\dagger$Corresponding author. Email: felixggh@if.usp.br}
}
\date{}

\AtEveryBibitem{\clearfield{month}}
\AtEveryBibitem{\clearlist{language}}
\AtEveryBibitem{\clearfield{note}}

\begin{document}
\maketitle
\setstretch{1.5}
\begin{sciabstract}
    Polar metals, materials that exhibit both electric polarization and high conductivity, can also host topological phases. Because free carriers strongly suppress distortive polar order and change the Fermi level, controlling charge dynamics is crucial for simultaneously tuning ferroelectric and topological phases in the same material. Here, we explore the experimental conditions that enable access to these phases in bismuth-doped Pb$_{1-x}$Sn$_x$Te epilayers. For samples in the topological phase at $x = 0.5$, we use terahertz time-domain spectroscopy to evaluate their complex permittivity as a function of temperature. We observe a non-monotonic variation in carrier concentration with bismuth doping, indicating a change in carrier type. By tracking the transverse optical phonon mode, we identify a ferroelectric phase transition when distortive polar order emerges below a critical temperature that depends on carrier concentration. We show that bismuth doping controls the metallicity-dependent order parameters in the softening and hardening phases. Our work demonstrates a tunable platform for engineering exotic states of matter that integrate metallicity, ferroelectricity and topology.
\end{sciabstract}

\clearpage
\section*{INTRODUCTION}

Hysteretic switching of permanent electrical polarization, first observed more than 100 years ago in crystals of Rochelle salt, revealed the phenomenon now known as ferroelectricity \autocite{PhysRev.17.475,ferroelect}. Later, metallicity---traditionally considered antagonistic to ferroelectricity---was incorporated into a new unified concept, giving rise to polar metals: ferroelectric-like materials with high electrical conductivity but without polarization reversal by an external field \autocite{AndersonBlount,firstpolarmetal,Zhou_2020}. 

A subclass of materials with polar order emerges from the breaking of inversion symmetry, evolving from a centrosymmetric to a polar crystal phase \autocite{PhysRevMaterials.7.010301}. One example of a ferroelectric phase arising from distortive polar order is the rhombohedral deformation of the rock salt structure in Pb$_{1-x}$Sn$_{x}$Te, which occurs below an $x$-dependent critical temperature ($T_\textrm{c}$). In its paraelectric phase, this ferroelectric instability is characterized by the softening of a transverse optical (TO) phonon mode. On the PbTe side of the alloy, incipient ferroelectricity has been observed in reflectivity spectroscopy \autocite{Burkhard:77}, inelastic x-ray scattering \autocite{NatComm7.12291}, neutron scattering measurements \autocite{DelaireetAl2011NM}, and, more recently, in terahertz time-domain spectroscopy (THz-TDS) experiments \autocite{PhysRevLett.128.075901,kawahala_thickness-dependent_2023,phononpolaritons}. For the intermediate range of $x$, the soft TO mode has also been reported with $T_\textrm{c}$ values from \qty{60}{K} to \qty{120}{K} \autocite{KAWAMURA1975341,okamura_terahertz_2022,hernandez_observation_2023}. On the other end, SnTe has shown signatures of a ferroelectric transition near \qty{100}{K} in resistance \autocite{KOBAYASHI1975875} and THz-TDS measurements \autocite{okamura_terahertz_2022}. In determining the polar axis, the polarization direction in SnTe was found to be along the [111] direction \autocite{PhysRevB.102.155132}, while in Pb$_{1-x}$Sn$_x$Te, the polar axis has been reported along both the [001] \autocite{ZhangPNAS} and [111] directions \autocite{PRL2025}, based on second-harmonic generation measurements.

Remarkably, increasing $x$ also leads to a topological transition from the trivial insulator PbTe to the topological crystalline insulator (TCI) SnTe \autocite{Hsieh_2012,Xu2012,assaf2016massive,ChengetAl2019PRL}. Pb$_{1-x}$Sn$_{x}$Te enters the topological phase for $x$ values above a critical point, where the gap closes and then reopens with inverted bands, which corresponds to $x > 0.32$ at low temperatures \autocite{PhysRevLett.16.1193,doi:10.1126/sciadv.1602510}. In this phase, the TO phonon modes acquire chirality in magnetic fields and exhibit properties that strongly depend on electronic contributions \autocite{hernandez_observation_2023}. In the absence of ferroelectric polarization, band inversion is expected to occur at all L points, leading to the above-mentioned TCI phase transition. In contrast, when ferroelectric order is established with polarization along the [111] axis, the valley degeneracy is lifted due to the breaking of spatial inversion symmetry. This symmetry breaking modifies the topological character of the material, resulting in the observation of the quantum Hall effect of surface states, due to the emergence of a topological insulator state protected by time-reversal symmetry \autocite{PRL2025}. Recently, thermoelectric measurements have also shown modulation of the topological states by the ferroelectric distortion \autocite{JACS2025}.

Nevertheless, increasing $x$ to access these ferroelectric and topological phases inevitably results in high free carrier concentrations in real samples. This occurs because as-grown films exhibit a p-type character due to native Sn vacancies, as reported in all the experimental studies mentioned above. The role of the sample's metallicity in the polar transition has been investigated in SnTe using electrical and x-ray techniques \autocite{PhysRevLett.37.772,PhysRevB.95.144101} and explored using terahertz driving fields in Pb$_{1-x}$Sn$_x$Te \autocite{sciadv_baldini}. In general, high carrier density on unintentionally p-doped samples was found to decrease $T_\textrm{c}$ for a fixed value of $x$. Although the dielectric response is large due to the large splitting between longitudinal and transversal phonon modes in this material, it is clear that a high carrier concentration limits the reach of spectroscopic methods due to enhancement of Drude's absorption, in particular observed for larger values of $x$ \autocite{okamura_terahertz_2022}.

To enable the establishment of the desired polar order, controlling the Fermi level through tuning of the Sn content and adjusting the carrier concentration via extrinsic doping is required. In this context, different dopants have been used to tune charge dynamics in this material system. For instance, indium has been employed to suppress metallic conduction \autocite{handa_terahertz_2024} or, on the contrary, to increase carrier density and induce superconductivity \autocite{PhysRevMaterials.5.094202}, and also to break inversion symmetry, enabling access to the Weyl semimetal phase \autocite{ZhangPNAS}. Additionally, bismuth (Bi) doping up to 0.9\% has been used to produce a giant Rashba splitting at the surface \autocite{volobuev_giant_2017}. However, a systematic study of p- and n-type extrinsic doping to demonstrate control over the order parameters of the polar transition within the topological phase has not been reported.

Here, we aim to explore the strong coupling between lattice dynamics and electronic topology that makes Pb$_{1-x}$Sn$_{x}$Te an ideal platform for investigating carrier concentration-dependent polar phenomena within a topological phase. Given that one approach to manufacturing polar metals is by doping polar insulators \autocite{PolarMetalsRev}, we investigate bismuth doping to tune the metallicity-driven distortive transition in epilayers with $x = 0.5$. We found that Bi doping enables a two-fold tunability of both carrier concentration and type, allowing precise control of the critical temperature for the polar transition over a broad range from \qty{65}{K} to \qty{95}{K}. Additionally, we determined the metallicity-dependent order parameters across the transition, revealing distinct critical behaviors and highlighting the influence of extrinsic doping on the nature of the distortive polar transition in this topological phase.

\section*{RESULTS}
\paragraph*{Doping level-dependent terahertz transmittance}\mbox{}\\
\indent We employed THz-TDS to measure the low-energy transmission spectra of Bi-doped Pb$_{0.5}$Sn$_{0.5}$Te films epitaxially grown on (111)-oriented Barium Fluoride (BaF$_2$) substrates. The measurement geometry is illustrated in Fig.~1a, with additional details of the experimental setup and procedures provided in the Methods section. The \qty{2}{\um}-thick epilayers investigated in this work were grown with varying Bi-doping levels, ranging from \qty{0}{\percent} (undoped) to \qty{0.15}{\percent}; see Fig.~1b. Further details on the growth process are also provided in Methods.

Figure~1c shows the room temperature ($T=\qty{300}{\K}$) THz transmittance spectra measured for three representative films: undoped, \qty{0.06}{\percent}, and \qty{0.15}{\percent} Bi-doped. Notably, the transmittance curves for the undoped and \qty{0.15}{\percent} Bi-doped samples (dark blue and red lines, respectively) are nearly identical, while the \qty{0.06}{\percent} Bi-doped film exhibits a transmittance (light blue line) approximately twice as high across the measured spectrum. This behavior indicates a non-monotonic dependence of the THz transmission on the Bi-doping level. Moreover, all curves exhibit an absorption feature centered near \qty{1}{\THz} (yellow circular areas), which agrees with the expected frequency of the TO phonon mode in Pb$_{1-x}$Sn$_x$Te systems \autocite{okamura_terahertz_2022,hernandez_observation_2023} at room temperature. The absorption peak is slightly red-shifted for the \qty{0.06}{\percent} Bi-doped film, suggesting a modification of the lattice dynamics compared to the other samples shown in Fig.~1c. Additionally, the low overall transmittance (below \qty{1}{\percent}) indicates the presence of free carrier screening effects, which play a significant role in the THz response of these films.

\vfill
\begin{figure}[t]
    \centering
    \includegraphics{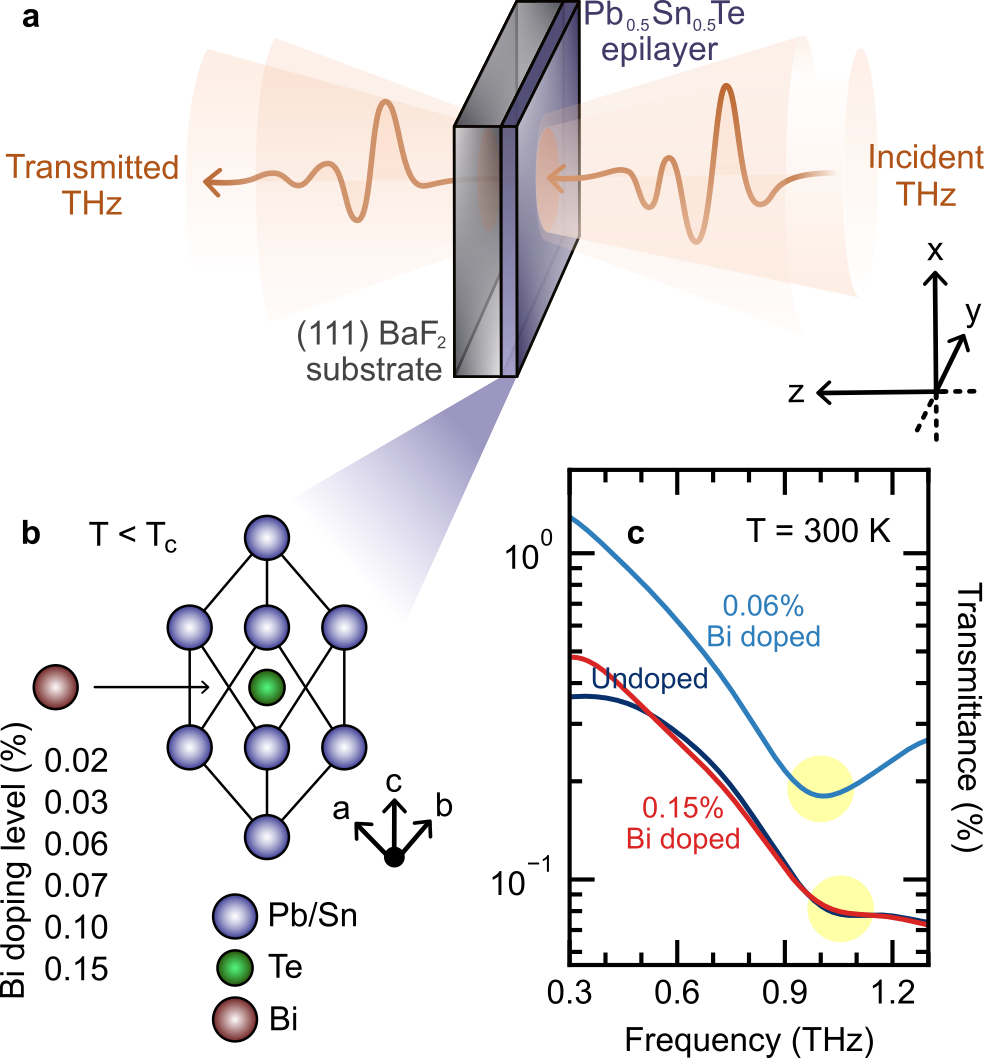}
    \caption{\textbf{THz transmittance of Pb$_{0.5}$Sn$_{0.5}$Te epilayers.} (\textbf{a})~Schematic diagram of the THz-TDS measurements, depicting the transmission of a THz beam through a sample (film-coated substrate). (\textbf{b})~Distorted cubic unit cell structure of the Bi-doped Pb$_{0.5}$Sn$_{0.5}$Te in the polar phase. (\textbf{c})~THz transmittance at \qty{300}{\K} for the films with \qty{0}{\percent} (dark blue curve), \qty{0.06}{\percent} (light blue curve), and \qty{0.15}{\percent} (red curve) Bi-doping levels.}
    \label{fig1}
\end{figure}
\vfill

\paragraph*{Temperature-dependent THz permittivity spectra}\mbox{}\\
\indent The frequency-dependent complex permittivity $\epsilon(\nu)=\epsilon_1(\nu)+i\epsilon_2(\nu)$ of films can be directly extracted from THz-TDS measurements using established transmission models; see Methods for details. Fig.~2 summarizes the real and imaginary components of the permittivity spectra obtained for the undoped, \qty{0.06}{\percent}, and \qty{0.15}{\percent} Bi-doped samples measured at temperatures ranging from \qty{300}{\K} (dark red lines) to \qty{10}{\K} (dark blue lines). Films with Bi-doping levels of \qty{0.02}{\percent}, \qty{0.03}{\percent}, \qty{0.07}{\percent}, and \qty{0.10}{\percent} were also measured, and their experimental permittivity curves are presented in the Supplementary Information.

\begin{figure}[H]
    \centering
    \includegraphics{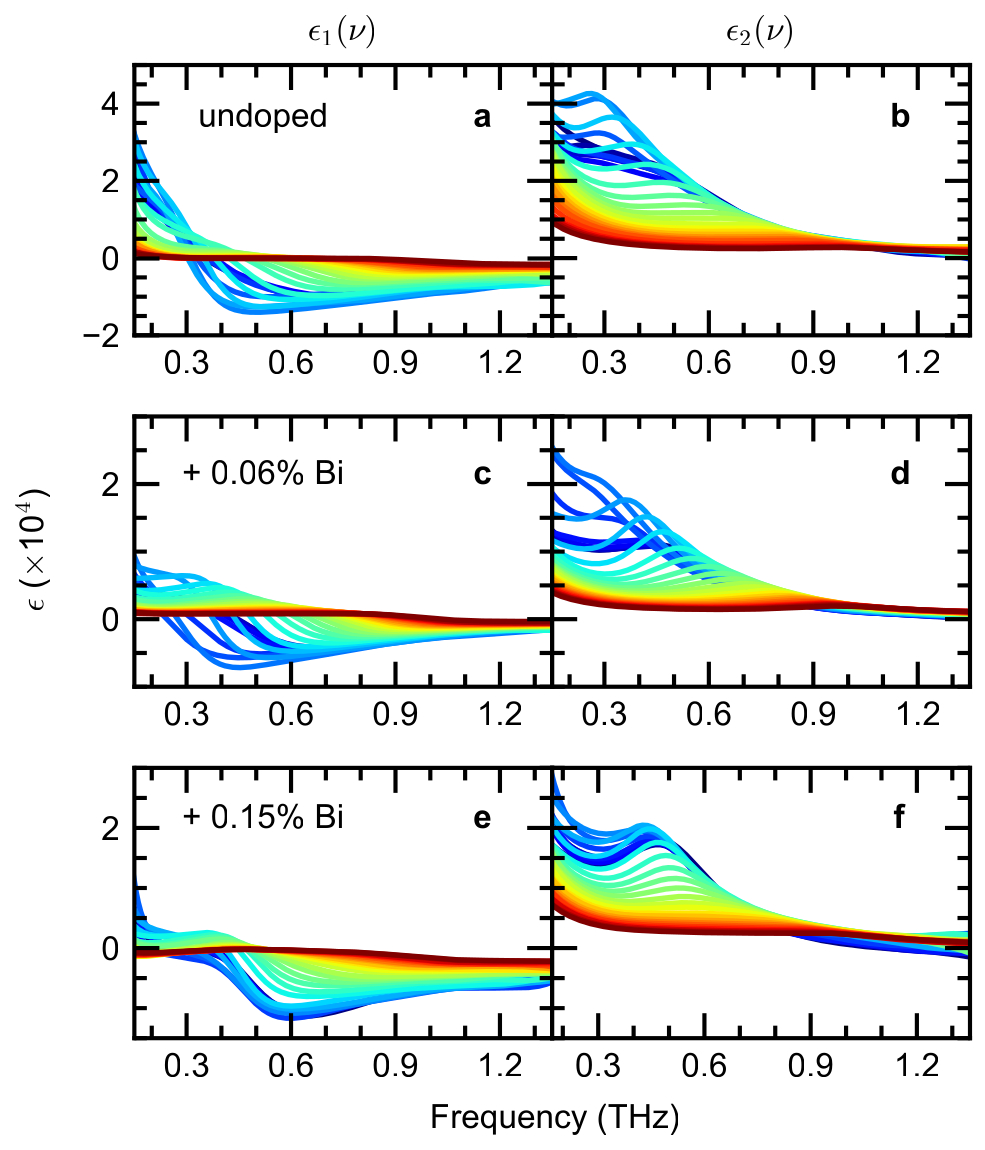}
    \caption{\textbf{Complex permittivity spectra.} (\textbf{a}, \textbf{c}, \textbf{e}) Real and (\textbf{b}, \textbf{d}, \textbf{f}) imaginary parts of the frequency-dependent permittivity spectra of the Pb$_{0.5}$Sn$_{0.5}$Te epilayers, extracted from the THz-TDS measurements at various temperatures, ranging from \qty{300}{\K} (dark red lines) to \qty{10}{\K} (dark blue lines) in approximately \qty{10}{\K} steps. The rows in the grid of the figure panels corresponds to the (\textbf{a}, \textbf{b}) undoped, (\textbf{c}, \textbf{d}) \qty{0.06}{\percent} Bi-doped, and (\textbf{e}, \textbf{f}) \qty{0.15}{\percent} Bi-doped samples.}
    \label{fig2}
\end{figure}

Across the measured frequency range, a prominent resonance feature is observed for all samples, sharpening and intensifying as the temperature decreases. This resonance shifts to lower frequencies with cooling, consistent with the softening of the TO phonon mode in the paraelectric phase due to ferroelectric instability. Below a critical temperature, however, the central frequency of the resonance begins to increase again, indicating a hardening of the TO mode in the ferroelectric phase. 

The resonance lineshape is highly asymmetric and deviates significantly from the pure Lorentzian profile typically associated with phonon contributions to the dielectric function. The additional spectral weight observed at low frequencies suggests the presence of free carrier effects, which modify the dielectric response, giving it a metallic behavior characterized by a local maximum of $\epsilon_2$ at zero frequency (Drude peak) \autocite{PhysRevMaterials.7.010301}. In addition, previous electrical measurements have shown an increase in resistance with temperature, indicating metallic behavior \autocite{MarcelosBiPbSnTe}. To disentangle the contributions of lattice and free carrier dynamics, we employ a combined Drude-Lorentz model, which treats the TO phonon as a damped Lorentz oscillator embedded within a Drude conductivity background \autocite{Burkhard:77,kawahala_thickness-dependent_2023}:
\begin{equation}\label{eq:drude-lorentz}
    \epsilon(\nu) = \epsilon_\infty + \frac{(\epsilon_\textrm{s}-\epsilon_\infty)\nu_\textrm{TO}^2}{\nu_\textrm{TO}^2-\nu^2-i\nu\Gamma_\textrm{TO}} - \frac{\nu_\textrm{p}^2}{\nu^2+i\nu\gamma},
\end{equation}
where $\epsilon_\textrm{s}$ and $\epsilon_\infty$ are the static and high-frequency dielectric constants, $\nu_\textrm{TO}$ and $\Gamma_\textrm{TO}$ are the TO phonon central frequency and linewidth, $\nu_\textrm{p}$ is the plasma frequency, and $\gamma$ is the carrier collision rate \autocite{fox_optical_2010}. This model captures both the phononic and electronic contributions to the dielectric response, allowing us to quantify their relative influence.

We used equation~\eqref{eq:drude-lorentz} to simultaneously fit the real and imaginary parts of the permittivity spectra for all measured temperatures and doping levels. During the optimization process, all model parameters were varied to achieve the best fit to the experimental data. As an example, Fig.~3 shows the fitted curves (black lines) for the \qty{0.06}{\percent} Bi-doped film at three representative temperatures: \qty{10}{\K} (blue circles), \qty{137}{\K} (green circles), and \qty{287}{\K} (red circles). The excellent agreement between the model and experimental data demonstrates the effectiveness of this approach in capturing the complex electrodynamic response of the Pb$_{0.5}$Sn$_{0.5}$Te films in the THz frequency range.

\begin{figure}[H]
    \centering
    \includegraphics{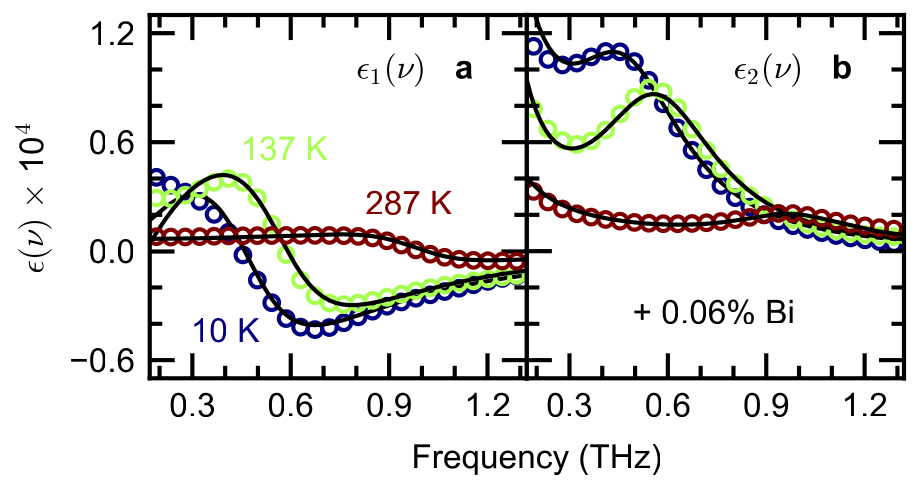}
    \caption{\textbf{Permittivity fits.} (\textbf{a}) Real and (\textbf{b}) imaginary parts of the complex permittivity data for the \qty{0.06}{\percent} Bi-doped film at \qty{10}{\K} (blue circles), \qty{137}{\K} (green circles), and \qty{287}{\K} (red circles). The black lines represent the optimal fits using equation~\eqref{eq:drude-lorentz}.}
    \label{fig3}
\end{figure}

\paragraph*{Bismuth doping-dependent metallicity}\mbox{}\\
\indent We begin by analyzing the Drude parameters of the permittivity fits, which describe the free carrier dynamics in the investigated films. These parameters provide critical insights into the interplay between carrier density, scattering mechanisms, and doping in Pb$_{0.5}$Sn$_{0.5}$Te. Fig.~4a shows the temperature dependence of the squared plasma frequency for the undoped (dark blue circles), \qty{0.06}{\percent} Bi-doped (light blue circles), and \qty{0.15}{\percent} Bi-doped films (red circles). This quantity is directly proportional to the carrier concentration $N$, as described by \autocite{Kittel}
\begin{equation}\label{eq:density}
    N = \frac{4\pi^2\epsilon_0m}{q^2}\nu_\textrm{p}^2,
\end{equation}
where $q$ is the carrier charge, $m=m^*m_0$ is the carrier mass, with $m^*$ and $m_0$ being the effective and free-electron mass, and $\epsilon_0$ is the vacuum permittivity. The data reveal two distinct regimes in the doping-dependent carrier concentration. For temperatures above approximately \qty{190}{\K}, $N$ decreases monotonically with increasing doping. In contrast, at lower temperatures, this dependence becomes non-monotonic, with the \qty{0.06}{\percent} Bi-doped sample exhibiting the lowest carrier concentration.

\begin{figure}[t]
    \centering
    \includegraphics{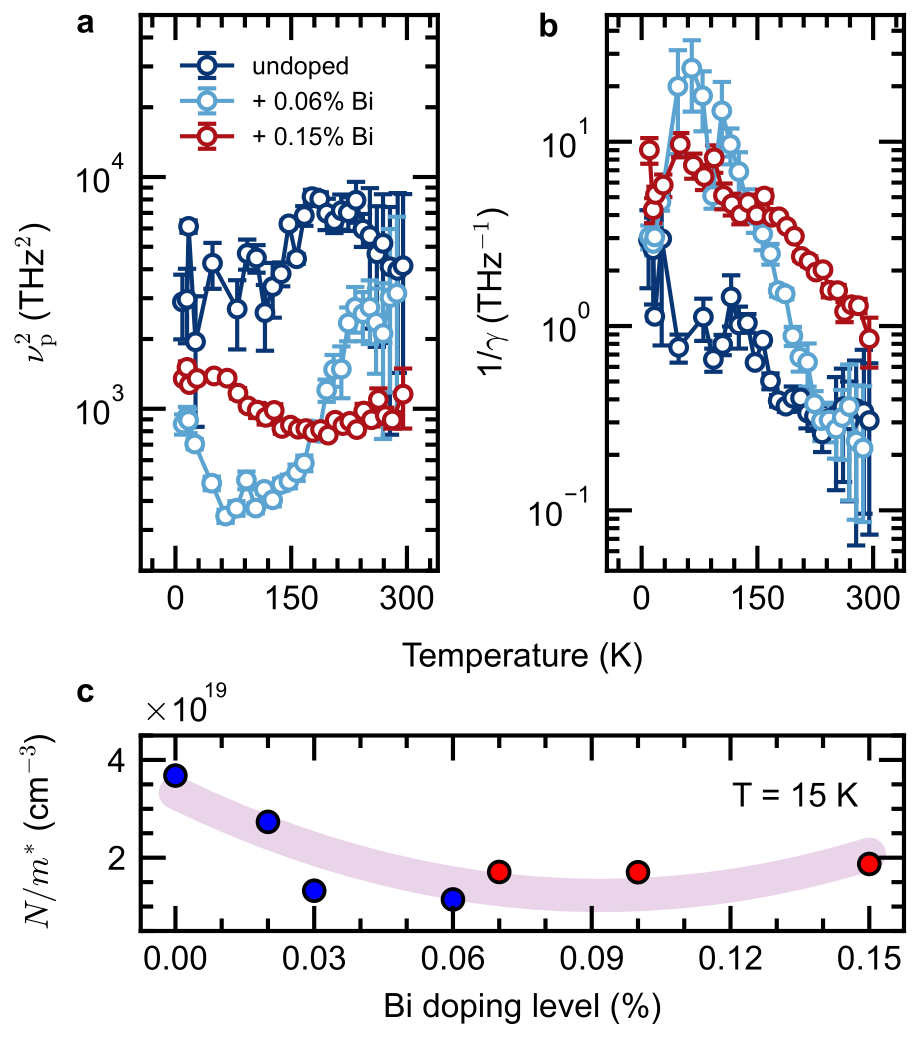}
    \caption{\textbf{Carrier dynamics}. Temperature dependence of the (\textbf{a}) squared plasma frequency and (\textbf{b}) inverse of the carrier collision rate for the undoped (dark blue circles), \qty{0.06}{\percent} Bi-doped (light blue circles), and \qty{0.15}{\percent} Bi-doped (red circles) films, calculated from the optimal fit parameters of equation~\eqref{eq:drude-lorentz} to the experimental permittivity data. (\textbf{c}) Ratio of the carrier concentration to its effective mass at a fixed temperature of \qty{15}{\K}, plotted as a function of the Bi-doping level. Error bars are smaller than the data point symbols. Blue and red circles represent the $p$- and $n$-type conductivity regimes, respectively. The shaded curve indicates the observed trend.}
    \label{fig4}
\end{figure}

The temperature dependence of $\nu_\textrm{p}^2$ exhibits distinct trends for each doping level, reflecting significant variations in carrier dynamics with doping. For the undoped sample, the behavior is consistent with carrier freeze-out, as we observe a reduction in the free carrier density with decreasing temperature. Conversely, for the \qty{0.15}{\percent} Bi-doped film, $N$ mainly increases with cooling, suggesting the emergence of impurity band conduction at high doping levels. The \qty{0.06}{\percent} Bi-doped sample exhibits an intermediate behavior, with a trend that reflects a competition between freeze-out and impurity band conduction. Additionally, the topological surface states may contribute to the low-temperature conductivity, further complicating the carrier dynamics.

Fig.~4b presents the inverse of the carrier collision rate as a function of temperature for the same doping levels. In all cases, $\gamma^{-1}$ mainly increases with cooling, indicating an enhancement in carrier mobility $\mu$, as these quantities are directly related through \autocite{Kittel}
\begin{equation}\label{eq:mobility}
    \mu = \frac{q}{m}\gamma^{-1}.
\end{equation}
In Fig.~4b, the undoped and \qty{0.15}{\percent} Bi-doped samples exhibit similar temperature dependencies, with $\gamma^{-1}$ increasing by approximately an order of magnitude as the temperature decreases from \qty{300}{\K} to \qty{10}{\K}. However, $\gamma^{-1}$ is roughly three times greater for the \qty{0.15}{\percent} Bi-doped film compared to the undoped sample at all temperatures, indicating a strong influence of doping on mobility. In contrast, the \qty{0.06}{\percent} Bi-doped sample shows a more pronounced temperature dependence, with $\gamma^{-1}$ increasing by around two orders of magnitude over the same temperature range. This behavior suggests an intermediate doping regime with enhanced sensitivity to temperature-dependent mobility changes.

We now turn to the control of the sample’s metallicity as a function of bismuth doping. Transport measurements on these samples revealed that the undoped film exhibits $p$-type conductivity \autocite{kawata_properties_2022}. As the Bi-doping level increases, the hole concentration decreases, with a transition to $n$-type conductivity occurring at a doping level of \qty{0.07}{\percent}. Beyond this point, the electron concentration increases with further doping. The sample reaches a carrier density similar to the undoped film, which is \qty{4.2e18}{\cm^{-3}} at low temperature, when the Bi-doping level is near \qty{0.10}{\percent}. This tuning of the carrier concentration was also found in our THz-TDS experiment.  Fig.4c shows the doping-level dependence of $N/m^*$ at \qty{15}{\K}, calculated using the $\nu_\textrm{p}^2$ data and equation~\eqref{eq:density}. Blue and red circles represent the $p$- and $n$-type conductivity regimes, respectively, as determined by transport measurements. 

By inserting the value of N from Hall effect measurements (see Supplementary Information) into the values for N/m* in Fig. 4c, our study suggests that the effective mass lies in the range of \num{0.1} to \num{0.2} for holes and electrons. Considering the anisotropy of the effective mass tensor and the low-temperature experimental values of its transverse component for holes of \num{0.039} in PbTe \autocite{PhysRev.135.A514} and \num{0.07} in SnTe \autocite{PhysRevB.98.195136}, we estimate $m^*_h = $ \num{0.084} and \num{0.14}, respectively.  As a first approximation, a linear interpolation for $x = 0.5$ yields $m^*_h =$ \num{0.11}, which agrees with our result. Although direct measurements of the electron effective mass $m^*_e$ in n-type SnTe are limited, its value in PbTe, $m^*_e =$ \num{0.20} at \qty{77}{K}, is higher than that for holes \autocite{PhysRev.135.A514} and may explain the observed asymmetry in Fig.~4c. The close agreement between our THz-TDS-derived carrier density and previous transport measurements highlights the consistency and reliability of our findings.

\paragraph*{Carrier concentration-dependent polar transition}\mbox{}\\
\indent The parameters extracted from the permittivity fits also offer important insights into the lattice dynamics of the measured films. Fig.~5a--g compare the temperature dependence of the TO phonon frequency (white circles, left axes) and the static dielectric constant (red squares, right axes) across all investigated doping levels. The background maps depict the imaginary part of the permittivity as a function of temperature and frequency.

\begin{figure}[p]
    \centering
    \includegraphics{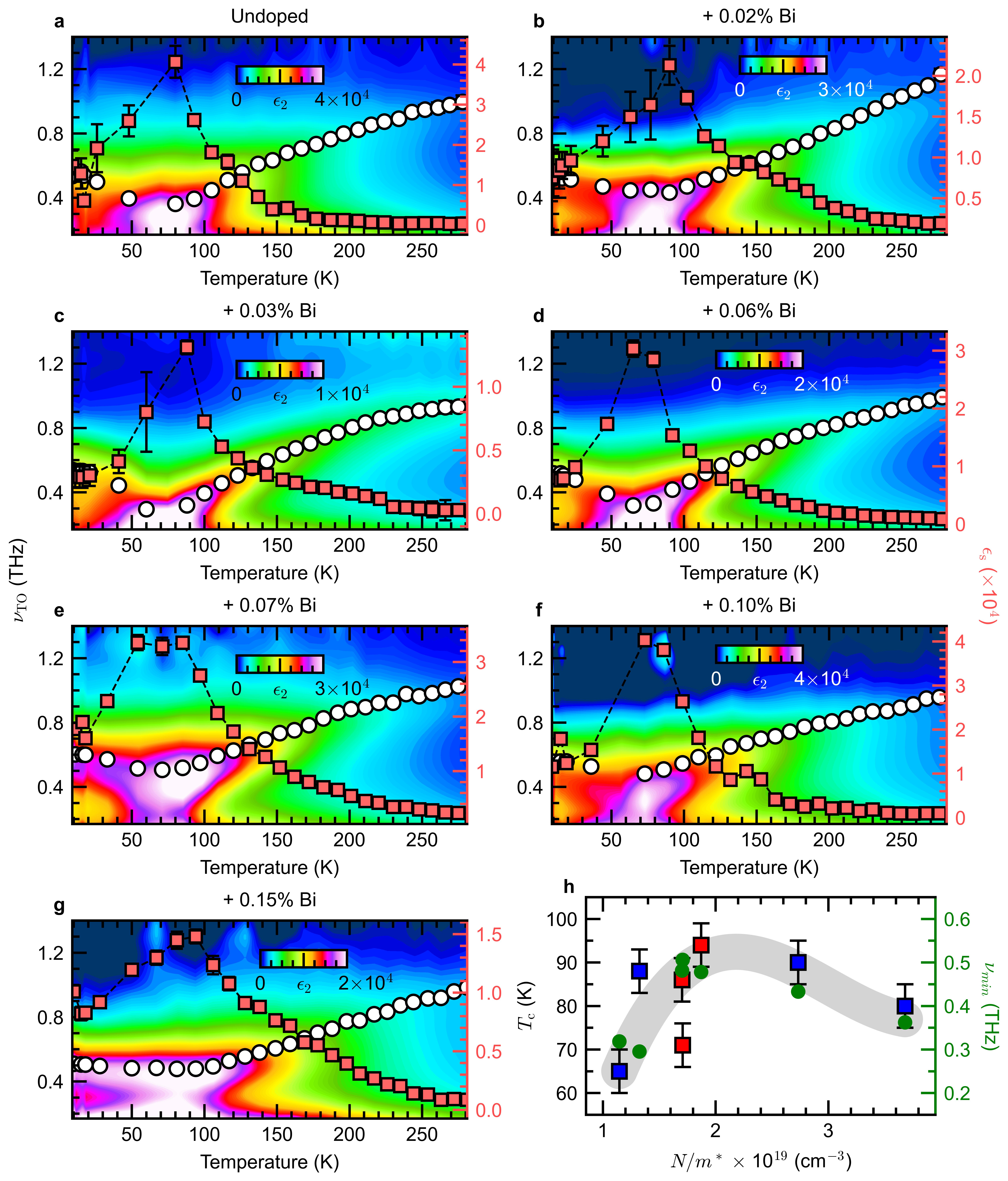}
    \caption{\textbf{Lattice dynamics.} (\textbf{a--g}) Temperature dependence of the TO phonon frequency (white circles, left axes) and static dielectric constant (red squares, right axes) for each investigated Bi-doping level, obtained from the optimal fit parameters of equation~\eqref{eq:drude-lorentz} to the experimental permittivity data. Background maps show the imaginary part of the permittivity as a function of temperature and frequency. (\textbf{h}) Carrier concentration dependence of the critical temperature (squares, left axis), and minimum value attained for the TO phonon frequency (green circles, right axis). Blue and red squares represent the $p$- and $n$-type conductivity regimes, respectively. The shaded curves indicate the observed trends.}
    \label{fig5}
\end{figure}

As each film is cooled from \qty{300}{\K}, $\nu_\textrm{TO}$ decreases while $\epsilon_\textrm{s}$ increases, reaching their respective minimum and maximum at a critical temperature $T_\textrm{c}$, below which the temperature dependence of $\nu_\textrm{TO}$ and $\epsilon_\textrm{s}$ becomes inverted. This critical temperature marks the onset of the ferroelectric phase transition. In thin films, epitaxial strain results in in-plane unfrozen modes and one out-of-plane frozen mode that drives the ferroelectric transition \autocite{okamura_terahertz_2022}. Although in-plane modes exhibit hardening that is weaker than the soft phonon polarized along the polar direction \autocite{RevModPhys.46.83}, THz-TDS allows us to determine $T_\textrm{c}$ by measuring the phonon mode that is polarized in-plane. Additionally, the maps reveal that the minima of $\nu_\textrm{TO}$ correlate with the maxima of $\epsilon_2$, further confirming the transition. The behavior in the temperature regime above $T_\textrm{c}$ is consistent with that expected for the paraelectric phase in a material exhibiting ferroelectric instability. Specifically, we observe $\epsilon_\textrm{s}$ to follow a Curie-Weiss law of the form \autocite{devonshire_theory_1954}
\begin{equation}\label{eq:es-curie}
    \epsilon_\textrm{s}(T) \propto (T-T_\textrm{c})^{-1}.
\end{equation}

\clearpage
Figure 5h shows the dependence of $T_\textrm{c}$ (squares, left axis) on the carrier concentration, which we extracted as the temperature corresponding to the maximum in $\epsilon_\textrm{s}$. Here, the nominal Bi-doping level of each sample is replaced by the experimentally determined $N/m^*$ value (see Fig.~4c). The same panel also displays the minimum TO phonon frequency (green circles, right axis) for each film, measured as $\nu_\textrm{min}=\nu_\textrm{TO}(T=T_\textrm{c})$. Interestingly, $T_\textrm{c}$ and $\nu_\textrm{min}$ exhibit an in-phase dependence on the carrier concentration, as indicated by the shaded gray trend curve. Contrary to the general assumption that increasing N should reduce $T_\textrm{c}$, we find that an optimal carrier concentration leads to a maximum in $T_\textrm{c}$ at intermediate doping levels, while $\nu_\textrm{min}$ reaches its minimum at the lowest concentrations. Remarkably, the film doped with \qty{0.03}{\percent} Bi shows the most promising parameters among the measured samples, combining a high ferroelectric transition temperature ($T_\textrm{c}\approx\qty{90}{\K}$), near-complete softening of the TO mode ($\nu_\textrm{min}<\qty{0.3}{\THz}$), and low carrier concentration ($N = \qty{2.2e18}{cm^{-3}}$).

The results in Fig.~5 systematically demonstrate that the ferroelectric transition is governed by both temperature and carrier concentration, revealing a strong coupling between dopant incorporation and lattice dynamics. This connection is particularly evident given that the ferroelectric order parameter is directly tied to the soft-mode frequency \autocite{RevModPhys.46.83}, enabling the assessment of Bi-doping effects through the temperature dependence of $\nu_\textrm{TO}$. For a continuous (second-order) displacive transition, $\nu_\textrm{TO}$ should vanish as $T$ approaches $T_\textrm{c}$ from either above or below, following the power law \autocite{cowley_structural_1980}
\begin{equation}\label{eq:landau}
    \nu_\textrm{TO} \propto \left|T-T_\textrm{c}\right|^\beta,
\end{equation}
where $\beta$ is the critical exponent of the order parameter. The second-order nature of the ferroelectric transition was confirmed in SnTe by inelastic neutron and x-ray scattering measurements at frequencies below 0.2 THz \autocite{newPawleyPRB,PhysRevB.95.144101}. Furthermore, 
similar $\Gamma$-point TO phonon behavior has been associated to the displacive nature of the transition in  Pb$_{1-x}$Sn$_{x}$Te \autocite{okamura_terahertz_2022,hernandez_observation_2023}. Notably, equation~\eqref{eq:landau} reduces to the Curie-Weiss law for $\nu_\textrm{TO}$ when $\beta=1/2$ (the mean-field value), consistent with the Lyddane-Sachs-Teller (LST) relation \autocite{Kittel,sievers_generalized_1991}, $\nu_\textrm{TO}^2\propto\epsilon_s^{-1}$; see equation~\eqref{eq:es-curie}.

To determine how the polarization fluctuations behave before ordering sets and how the ferroelectric phase emerges, we extracted the critical exponent $\beta$ from a log-log scaling analysis of $\nu_\textrm{TO}$ across the transition. Fig.~6 shows the $\log(\nu_\textrm{TO})$ plotted against (a) $-\log[(T_\textrm{c}-T)/T_\textrm{c}]$ for the hardening ferroelectric phase ($T<T_\textrm{c}$), and (b) $\log[(T-T_\textrm{c})/T_\textrm{c}]$ for the softening paraelectric phase ($T>T_\textrm{c}$), with vertical offsets applied for clarity. As predicted by equation~\eqref{eq:landau}, the slopes of these plots should directly yield $\beta$. 

\begin{figure}[h]
    \centering
    \includegraphics{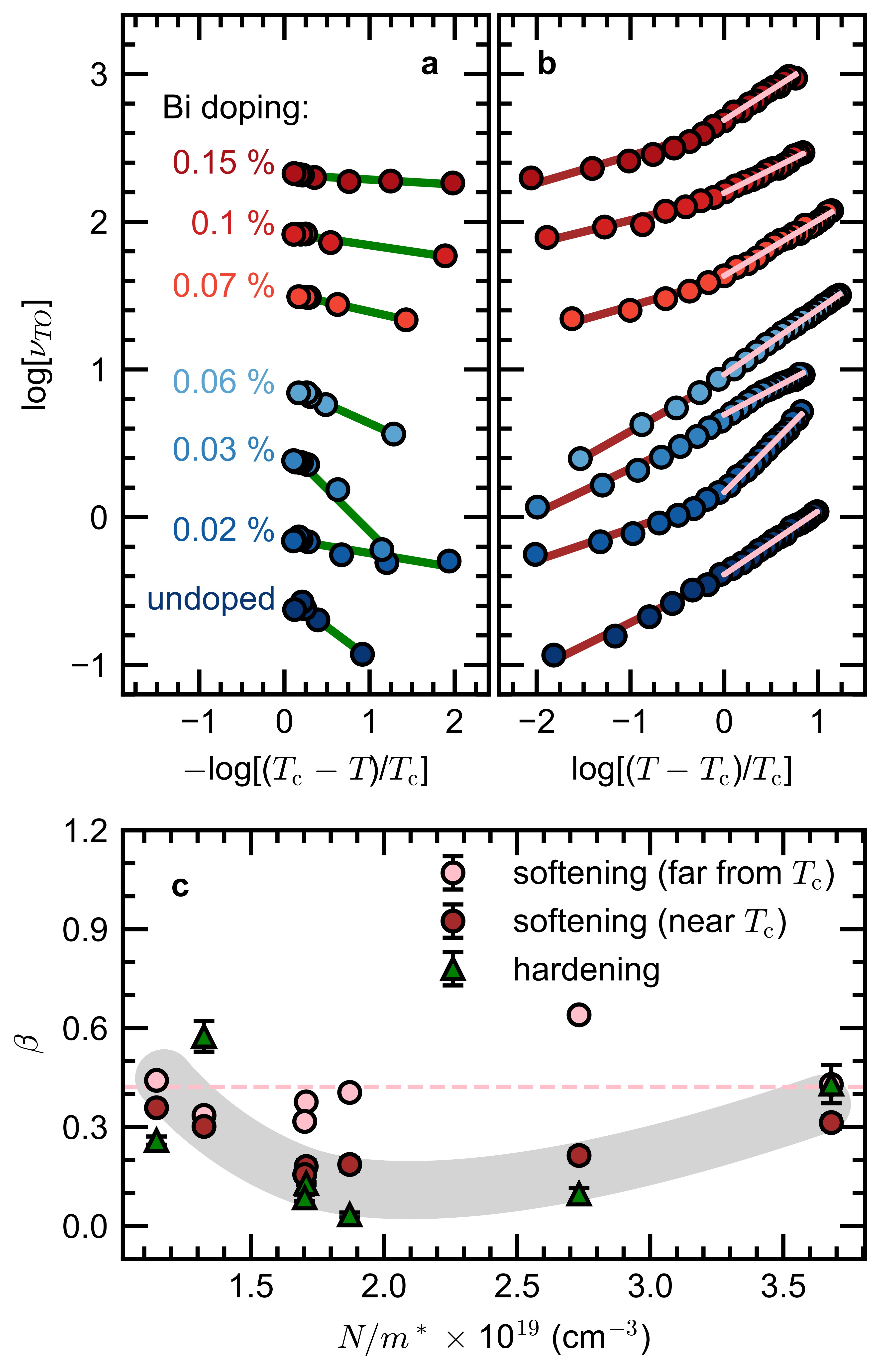}
    \caption{\textbf{Critical exponents of the order parameter.} Log-log plots of the TO phonon frequency versus reduced temperature for the (\textbf{a}) hardening and (\textbf{b}) softening phases of the polar transition, with data for different Bi-doping levels vertically offset for clarity. Solid lines show linear fits to three distinct temperature regimes: hardening phase (green), softening phase near transition (dark-red), and softening phase far from transition (light-red). (\textbf{c}) Corresponding critical exponents extracted from the slope of these fits as a function of the film's carrier concentration. The shaded gray curve highlights the trend observed for both hardening and softening-near-$T_\textrm{c}$ regimes. The dashed horizontal line marks the average $\beta$ value determined from the softening far-from-$T_\textrm{c}$ data.}
    \label{fig6}
\end{figure}

The analysis reveals distinct critical behaviors: while the hardening phase exhibits a single scaling regime, the softening phase shows two distinct regions---one near ($1<T/T_\textrm{c}<2$), and another far ($T>2T_\textrm{c}$) from $T_\textrm{c}$---displaying modified exponents. This temperature-dependent scaling indicates a crossover from fluctuation-dominated dynamics near $T_\textrm{c}$ to mean-field-like behavior at elevated temperatures, as confirmed by the linear fits shown in Fig.~6a,b (solid lines). The extracted $\beta$ values for all three scaling regions are systematically presented in Fig.~6c as a function of the experimentally determined $N/m^*$.

Across the investigated doping levels, the critical exponent for the high-temperature softening regime (light-red circles) exhibits a consistent average value of $\beta=\num{0.42\pm0.04}$---as indicated by the dashed horizontal line---which is significantly close to the classical mean-field prediction ($\beta=\num{0.5}$). In contrast, near $T_\textrm{c}$ in the softening phase (dark-red circles), $\beta<0.4$ for all investigated carrier concentrations, reflecting stronger polarization fluctuation effects. In particular, $\beta$ reaches a minimum at intermediate $N/m^*$ values, as emphasized by the gray trend curve.

The critical exponent $\beta$ for the hardening phase (green triangles) exhibits a carrier concentration dependence similar to that of the softening regime near $T_\textrm{c}$, though with slightly lower values overall. Remarkably, the sample with the Bi doping of \qty{0.03}{\percent} shows the steepest hardening, with a critical exponent value approximately twice that observed in the softening phase. Weak hardening is observed throughout the n-type doping range. Similar behavior, inconsistent with the prediction of Landau theory for ferroelectrics \autocite{Kittel,PhysRevB.95.144101}, has been reported in materials with degenerate soft phonons \autocite{RevModPhys.46.83}. Thus, it suggests that high Bi doping strongly influences the symmetry breaking involved in the polar phase transition. Despite this variability, the average values of $\beta$ for the ferroelectric and critical paraelectric regimes are consistent within uncertainty: \num{0.23\pm0.08} versus \num{0.24\pm0.03}, respectively. 

\section*{DISCUSSION}

Our temperature-dependent THz-TDS experiments provide an effective approach for probing the dielectric response of Bi-doped Pb$_{0.5}$Sn$_{0.5}$Te topological epilayers, offering robust characterization of their metallicity and polar order dynamics. Through Drude-Lorentz modeling of the THz permittivity spectra, we quantitatively decoupled the TO phonon contribution from the electronic background. This reveals a rich interplay between free carriers and lattice dynamics driven by Bi doping, shedding light on the fundamental mechanisms governing the polar transition in this topological material.

The carrier concentration exhibits a non-monotonic dependence on Bi doping, reaching a minimum at \qty{0.06}{\percent} Bi that aligns with the $p$- to $n$-type transition. This critical doping threshold demonstrates that Bi acts as an extrinsic dopant that compensates intrinsic metal vacancies in Pb$_{0.5}$Sn$_{0.5}$Te, enabling controlled modulation of electronic properties. The observed two-fold variation in carrier concentration quantitatively matches values obtained from electronic transport measurements, confirming the consistency between THz spectroscopy and transport characterization. From this analysis, we extract effective masses for holes and electrons that agree well with prior estimates. However, direct cyclotron resonance measurements via magneto-THz-TDS remain essential to refine these parameters for this material family. 

Across all doping levels, the temperature dependence of the TO phonon reveals a well-defined ferroelectric transition, characterized by mode softening in the paraelectric phase and hardening below a finite critical temperature $T_\textrm{c}$. This transition is unambiguously identified via peaks in the static dielectric constant---a hallmark of polar order. Notably, carrier concentration profoundly influences lattice dynamics, with the paraelectric phase showing enhanced temperature sensitivity compared to the ferroelectric phase. These findings demonstrate strong carrier-polar distortion coupling, underscoring the decisive role of Bi doping in tuning both electronic and structural properties.

We demonstrate the doping-driven tunability of $T_\textrm{c}$ over a broad range from \qty{65}{\K} to \qty{95}{\K}, values that align closely with our previous work on Pb$_{1-x}$Sn$_{x}$Te epilayers with varying Sn content \autocite{hernandez_observation_2023}. In particular, despite the known influence of substrates on defect formation in epitaxial systems, the value of $T_\textrm{c}=\qty{80}{\K}$ obtained for the undoped sample shows striking consistency with prior studies on Pb$_{1-x}$Sn$_{x}$Te grown on different substrates---including In-free \autocite{okamura_terahertz_2022} and In-doped \autocite{handa_terahertz_2024} films grown on InP, as well as undoped films grown on GaAs \autocite{sciadv_baldini}.

We emphasize that $T_\textrm{c}$ alone cannot fully characterize the polar transition. Through Landau theory modeling of the order parameter, we systematically evaluate the evolution of the critical exponent across temperature and doping regimes. While we found that intermediate carrier concentration increases $T_\textrm{c}$, it significantly reduces the critical exponent of the hardening phase---revealing its dual role in enhancing the transition temperature while fundamentally modifying the nature of the transition. This analysis is essential for quantifying polarization fluctuations, which strongly interact with charge carriers and suppress further softening of the TO phonon mode. Together, these observations underscore the complexity of ferroelectric behavior in this topological system, where conventional metrics may fail to capture the full evolution of the order parameter.

In summary, our results establish extrinsic doping as a powerful control parameter for simultaneously tuning metallicity and ferroelectric properties in polar topological epilayers. The observed coupling between carrier concentration and lattice dynamics highlights this material system as an ideal platform for fundamental studies of polar order in topological materials. These findings open new avenues for designing quantum materials with coupled electronic and structural responses.

\section*{METHODS}
\paragraph*{Samples}\mbox{}\\
\indent Pb$_{0.5}$Sn$_{0.5}$Te epitaxial films used for the investigations were grown by molecular beam epitaxy (MBE) using a Riber 32P MBE system on thin slices ($<\qty{1}{\mm}$) of BaF$_2$ substrates cleaved along the (111) crystalline plane \autocite{rappl_molecular_1998,kawata_properties_2022,Okazakithesis}. BaF$_2$ is transparent to radiation ranging from ultraviolet (\qty{0.2}{\um}, in wavelength) to infrared (\qty{10}{\um}), exhibiting a good transmittance in the THz band (around \qty{300}{\um}) \autocite{kaplunov_optical_2021}. The lattice mismatch between the film and substrate is of \qty{3}{\percent}, and their linear thermal expansion coefficients are very similar. These properties make BaF$_2$ an ideal substrate for the epitaxial growth of Pb$_{0.5}$Sn$_{0.5}$Te, as well as for its optical characterization at low temperatures. During the growth, three effusion cells charged with solid sources of PbTe, SnTe, and Bi$_2$Te$_3$ were used \autocite{kawata_properties_2022,Biancathesis}. Equal beam equivalent pressures (BEPs) for PbTe and SnTe were used to ensure a nominal Sn content of 0.5, which was verified by measuring the lattice constant of the films, resulting in $x=(0.51\pm0.02$). Growth dynamics (monitored by RHEED) showed initial island nucleation, followed by coalescence into a continuous film and transition to step-flow growth after $\sim$ \qty{60}{nm}. The substrate was held at \qty{240}{\celsius}, yielding a deposition rate of \qty{0.25}{nm/s} and a final film thickness of approximately \qty{2}{\um}. The bismuth doping was controlled by adjusting the Bi$_2$Te$_3$ cell temperature, with the nominal Bi concentration estimated from the ratio of its BEP to the sum of PbTe and SnTe BEPs. The temperature of the Bi$_2$Te$_3$ source was varied to obtain a series of Bi-doped Pb$_{0.5}$Sn$_{0.5}$Te samples with the following doping levels: \qty{0}{\percent} (undoped), \qty{0.02}{\percent}, \qty{0.03}{\percent}, \qty{0.06}{\percent}, \qty{0.07}{\percent}, \qty{0.10}{\percent}, and \qty{0.15}{\percent}. The observed change from p- to n-type doping is attributed to the valence configurations of the constituent elements. Pb and Sn have 2+ valence states, while Te is 2−. For the intrinsic material, the metal ions share their valence electrons with tellurium, forming the ionic crystal. When Bi (valence 3+) is introduced substitutionally at Pb/Sn sites, it shares two electrons in the ionic crystal and contributes an extra electron. As the doping level rises, the hole concentration reduces, resulting in effective extrinsic n-type doping of the Pb$_{0.5}$Sn$_{0.5}$Te crystal. At low temperature, electrical characterization of the sample series revealed carrier mobility in the range of \num{2.6e3} to \qty{5.8e3}{cm^2/\V\s}, and carrier concentration ranging from \num{2.2} to \qty{5.8e18}{cm^{-3}} (see Supplementary Information). The films have been quantitatively characterized by high-resolution X-ray diffraction, reciprocal space mapping, and ARPES showing fully relaxed, high-quality films where Bi doping has no influence on the structural properties\autocite{kawata_properties_2022,MarcelosBiPbSnTe}.

\paragraph*{Measurements}\mbox{}\\
\indent The samples were studied by temperature-dependent THz-TDS measurements using a conventional setup in a transmission geometry \autocite{kawahala_thickness-dependent_2023,kawahala_shaping_2025}. Broadband THz pulses were generated with a biased photoconductive antenna (PCA) pumped by infrared (IR) pulses from a mode-locked Ti:sapphire laser oscillator tuned to \qty{780}{\nm}, with a pulse duration of \qty{130}{\fs} and a repetition rate of \qty{76}{\MHz}. Each sample was mounted inside a cold finger cryocooler equipped with polytetrafluoroethylene windows, aligned with the THz beam path, allowing measurements from \qty{300}{\K} to \qty{10}{\K}. Off-axis parabolic mirrors were used to focus the THz beam onto the sample's film-coated face with a \qty{3}{\mm} spot diameter and to collect the transmitted beam leaving the back of the substrate \autocite{Dias2025}. The transmission was probed by another optically gated PCA, enabling time-domain recordings of the THz electric field. By Fourier transforming the gathered time-domain data, both the amplitude and phase components of the frequency-dependent THz electric fields were obtained. As these quantities carry information about the entire transmission system, we isolate the film contribution by determining the complex transmission coefficient
\begin{equation}
    \mathcal{T}(\nu) = \frac{E_\textrm{sam}(\nu)}{E_\textrm{ref}(\nu)}\exp\left[\frac{2\pi i\nu}{c}(1-n)\Delta L\right],
\end{equation}
where $E_\textrm{sam}, E_\textrm{ref}$ are the frequency-dependent sample (film over substrate) and reference (uncoated substrate) electric fields, $c$ is the speed of light in vacuum, $n$ is the substrate refractive index, and $\Delta L$ is the small thickness difference between the sample and reference substrates. The THz transmittance is computed as the square of the transmission coefficient amplitude $\left|\mathcal{T}(\nu)\right|^2$. Furthermore, the complex permittivity spectrum $\epsilon(\nu)=\epsilon_1(\nu)+i\epsilon_2(\nu)$ of a film with thickness $d$ can be determined using conventional transmission models in the thin film limit \autocite{lloyd-hughes_review_2012,neu_tutorial_2018} with appropriated windowing \autocite{Marulanda2025}
\begin{equation}
    \epsilon(\nu) = i(1+n)\left[\frac{1}{\mathcal{T}(\nu)}-1\right]\frac{c}{2\pi\nu d} - n.
\end{equation}

\noindent\textbf{Data Availability:} All data generated or analyzed during this study are included in this published article and its supplementary information file.\\
\\
\noindent\textbf{Acknowledgments:} This work was supported by the São Paulo Research Foundation (FAPESP), Grants Nos. 2021/12470-8 and 2023/04245-0.  F.G.G.H. acknowledges financial support from Grant No. 306550/2023-7 of the National Council for Scientific and Technological Development (CNPq). E.D.S. acknowledges support from CNPq Grant No. 407815/2022-8. N.M.K. acknowledges support from FAPESP Grant No. 2023/11158-6. P. H.O.R. and E.A. acknowledge support from CNPq Nos. 307192/2021-0 and 302288/2022-8, respectively.\\
\\
\noindent\textbf{Author Contributions:} F.G.G.H. conceived the project. B.A.K., P.H.O.R., and E.A. grew the samples. N.M.K. built the terahertz spectrometer and assisted E.D.S. with the measurements. N.M.K. and E.D.S. analyzed the experimental data. N.M.K., E.D.S. and F.G.G.H. prepared the manuscript. F.G.G.H. supervised the project. All authors discussed the results and commented on the manuscript.\\
\\
\noindent\textbf{Competing Interests:} All authors declare no financial or non-financial competing interests.\\

\setstretch{1}
\printbibliography

\clearpage

\renewcommand{\thepage}{S\arabic{page}}  
\renewcommand{\thesection}{S\arabic{section}}   
\renewcommand{\thetable}{S\arabic{table}}   
\renewcommand{\thefigure}{S\arabic{figure}}
\renewcommand{\theequation}{S\arabic{equation}}

\renewcommand{\figurename}{Fig.}
\renewcommand{\tablename}{Table}

\setcounter{equation}{0}
\setcounter{figure}{0}
\section*{Supplementary Information}

\begin{figure}[H]
    \centering
    \includegraphics{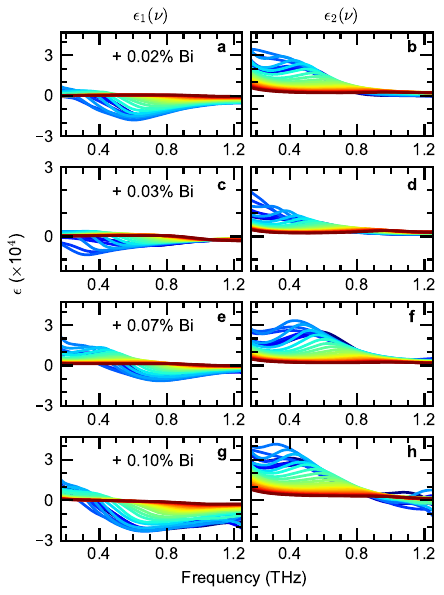}
    \caption{\textbf{Complex permittivity spectra.} (\textbf{a}, \textbf{c}, \textbf{e}, \textbf{g}) Real and (\textbf{b}, \textbf{d}, \textbf{f}, \textbf{h}) imaginary parts of the frequency-dependent permittivity spectra of the Pb$_{0.5}$Sn$_{0.5}$Te epifilms, extracted from the THz-TDS measurements for several temperatures from \qty{300}{\K} (dark red lines) to \qty{10}{\K} (dark blue lines) in steps of approximately \qty{10}{\K}. The rows in the grid of figure panels corresponds to the (\textbf{a}, \textbf{b}) \qty{0.02}{\percent}, (\textbf{c}, \textbf{d}) \qty{0.03}{\percent}, (\textbf{e}, \textbf{f}) \qty{0.07}{\percent} Bi-doped and (\textbf{g}, \textbf{h}) \qty{0.10}{\percent} Bi-doped samples.}
    \label{fig1_SI}
\end{figure}

\begin{table}[h!]
\centering
\caption{Data of Bi-doped Pb$_{0.5}$Sn$_{0.5}$Te epitaxial films grown by MBE on (111) BaF$_2$. Parameters include Bi$_2$Te$_3$ source temperature ($T$), beam equivalent pressure (BEP), nominal Bi content ($x_{\text{Bi}}$), carrier type and concentration ($N$), resistivity ($\rho$), and Hall mobility ($\mu$) measured at 13 K. Values extracted from [37].}
\label{tab:bi_te_properties}

\begin{tabular}{|c|c|c|c|c|c|c|}
\hline
\multicolumn{3}{|c|}{\textbf{Parameters of Bi$_2$Te$_3$}} & \multicolumn{4}{c|}{\textbf{Electrical properties at 13 K}} \\
\hline
\textbf{$T$ (°C)} & \textbf{BEP ($10^{-10}$ Torr)} & \textbf{$x_\text{Bi}$} & \textbf{Type} & \textbf{$\rho$ ($10^{-4}\,\Omega$cm)} & \textbf{$N$ ($10^{18}$ cm$^{-3}$)} & \textbf{$\mu$ ($10^{3}$ cm$^2$/Vs)} \\
\hline
\dots & \dots & 0.00 & p & 2.7 & 4.2 & 5.4 \\
355 & 3.0 & 0.02 & p & 2.9 & 3.7 & 5.8 \\
380 & 5.0 & 0.03 & p & 7.3 & 2.2 & 3.8 \\
393 & 9.0 & 0.06 & p & 8.2 & 2.9 & 2.6 \\
399 & 11.0 & 0.07 & n & 6.6 & 3.2 & 2.9 \\
393 & 16.0 & 0.10 & n & 2.8 & 4.4 & 5.1 \\
413 & 23.0 & 0.15 & n & 3.0 & 5.8 & 3.6 \\
\hline
\end{tabular}
\end{table}

\end{document}